\documentclass[preprint,12pt,3p]{elsarticle}

\usepackage{matlab-prettifier}
\usepackage{graphicx}
\usepackage{makecell}
\usepackage{xfrac}
\usepackage{parskip}
\usepackage{adjustbox}
\usepackage{setspace}
\usepackage{lscape}
\usepackage[backref=page]{hyperref}
\hypersetup{ 
backref=true, 
bookmarks=true, 
pdftoolbar=true, 
pdfmenubar=true, 
pdffitwindow=false, 
pdfstartview={FitH}, 
pdftitle={}, 
pdfauthor={}, 
pdfsubject={}, 
pdfkeywords={}{}, 
pdfnewwindow=true, 
colorlinks=true, 
linkcolor=BlueViolet, 
citecolor=maroon, 
urlcolor=blue,
}
\usepackage{comment}
\usepackage{subfig}
\usepackage{caption}
\usepackage{longtable}
\usepackage{amssymb}
\usepackage{amsmath}
\usepackage{gensymb}
\usepackage{siunitx}

\usepackage{natbib}
\biboptions{comma,round}
\setcitestyle{authoryear}

\begin{document}
\begin{frontmatter}

\title{Trends and biases in the social cost of carbon}

\author[label1,label2,label3,label4,label5,label6]{Richard S.J. Tol\corref{cor1}
}
\address[label1]{Department of Economics, University of Sussex, Falmer, United Kingdom}
\address[label2]{Institute for Environmental Studies, Vrije Universiteit, Amsterdam, The Netherlands}
\address[label3]{Department of Spatial Economics, Vrije Universiteit, Amsterdam, The Netherlands}
\address[label4]{Tinbergen Institute, Amsterdam, The Netherlands}
\address[label5]{CESifo, Munich, Germany}
\address[label6]{Payne Institute for Public Policy, Colorado School of Mines, Golden, CO, USA}

\cortext[cor1]{Jubilee Building, BN1 9SL, UK}

\ead{r.tol@sussex.ac.uk}
\ead[url]{http://www.ae-info.org/ae/Member/Tol\_Richard}

\begin{abstract}
An updated and extended meta-analysis confirms that the central estimate of the social cost of carbon is around \$200/tC with a large, right-skewed uncertainty and trending up. The pure rate of time preference and the inverse of the elasticity of intertemporal substitution are key assumptions, the total impact of 2.5\celsius warming less so. The social cost of carbon is much higher if climate change is assumed to affect economic growth rather than the level of output and welfare. The literature is dominated by a relatively small network of authors, based in a few countries. Publication and citation bias have pushed the social cost of carbon up.
\end{abstract}

\end{frontmatter}

\section{Introduction}
The social cost of carbon is a central statistic in climate policy. It measures the benefit of slightly reducing carbon dioxide emissions, justifying (or not) policies to do that \citep{Greenstone2013, Pizer2014}. There is therefore a large literature on the social cost of carbon, estimating its size, critiquing estimates, and commenting on its (lack of) application \citep{NAS2017, Pezzey2019}. This paper contributes by updating a previous meta-analysis \citep{Tol2023NCC} and exploring fields recently added to the meta-database, coding the assumptions underlying estimates of the social cost of carbon. Furthermore, I study citation and co-author networks and their influence on published estimates, showing signs of both publication and citation bias.

Previous meta-analyses focused on the characterization of the uncertainty about the social cost of carbon \citep{Tol2005, Tol2011}, the detection of trends \citep{Wang2019, Tol2023NCC}, and the discovery of publication bias \citep{Havranek2015, Tol2018REEP}. I briefly revisit uncertainty and trends as nothing much has changed: The uncertainty about the social cost of carbon is right-skewed and large, with a thick, perhaps fat fail \citep{Anthoff2022}. Estimates of the social cost of carbon have increased over time. The analysis below re-affirms these findings using a larger, more refined database.

I shed new light on publication bias. \citet{Havranek2015} use tests for publication bias that are appropriate for regression results\textemdash estimates of the social cost of carbon are not. Instead, I follow \citet{Tol2018REEP}, testing for confirmation bias\textemdash do later estimates agree or disagree with earlier ones? I improve on that paper in two ways. The newly documented citation network allows me to distinguish between earlier estimates that are surely known to the researchers and those that may not have been. A richer set of assumptions underlying the estimates of the social cost of carbon allows me to test whether researchers were influenced by the \emph{results} of previous studies or by their \emph{assumptions}.

Furthermore, this paper tests for citation bias in the social cost of carbon literature. Citation bias has been studied in other literatures \citep{Nieminen2007, Radicchi2012, Jannot2013}, but not for the social cost of carbon. As with publication bias, tests for citation bias focus on significance and are therefore not applicable here. Instead, I test on effect size and underlying assumptions.

The paper proceeds as follows. Section \ref{sc:data} discusses the descriptive statistics of the extended database, setting up the hypothesis to be tested in Section \ref{sc:analysis}. Section \ref{sc:conclude} discusses the results and their implications.

\section{Data and descriptive statistics}
\label{sc:data}
The data has been collected over two decades, starting with the meta-analysis of \citet{Tol2005}. The current database extends the one used by \citet{Tol2023NCC} with more papers and more estimates. The updated data also contains more information about the assumed intertemporal welfare functions and the impacts of climate change. Furthermore, I added fields on the authors and their affiliations, as well as on citations. The data is described in \citet{Tol2023meta}. The database is on \href{https://github.com/rtol/metascc/tree/master}{GitHub}.

\begin{figure}
    \centering
    \includegraphics[width = \textwidth]{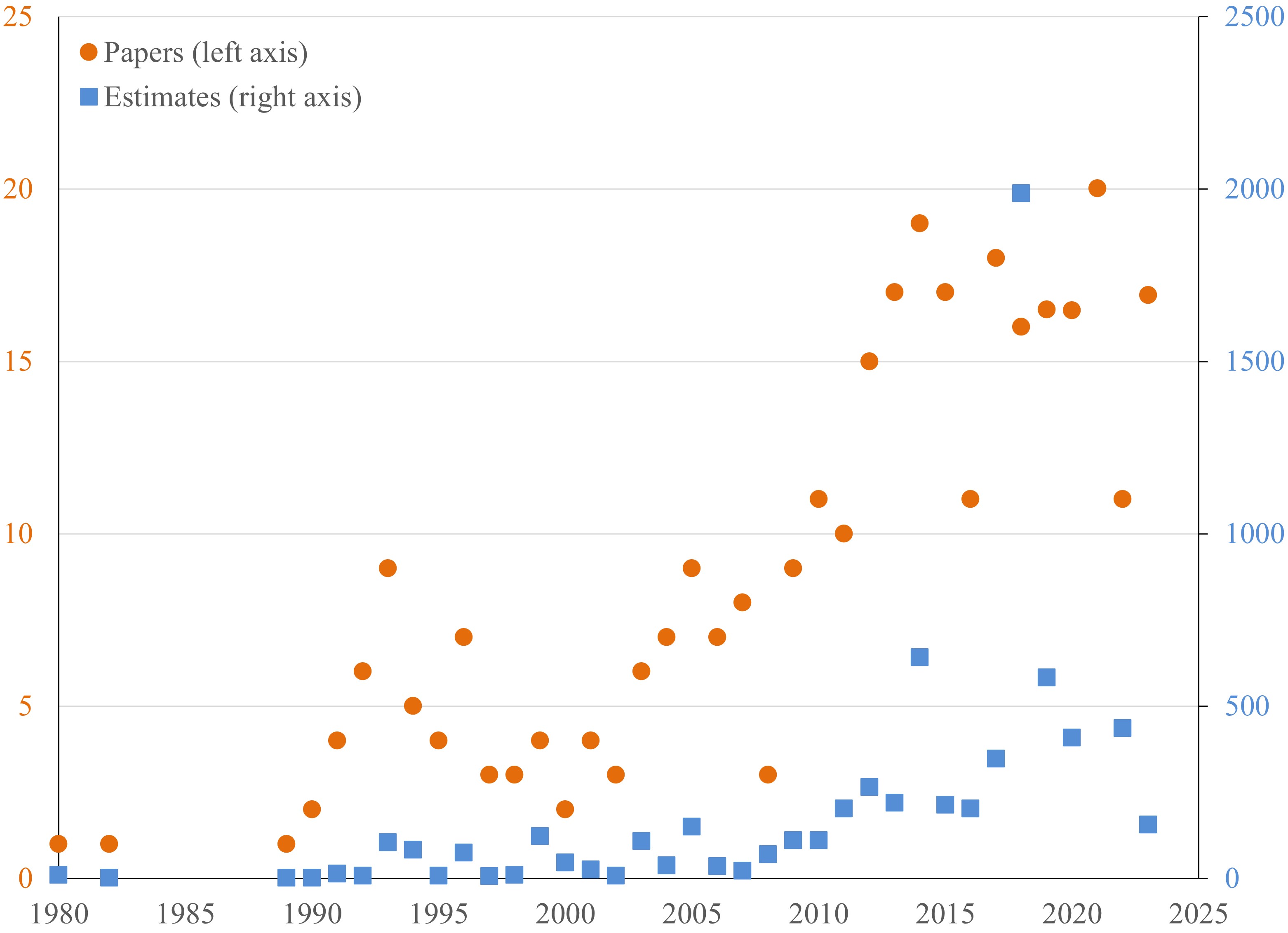}
    \caption{The number of papers on and estimates of the social cost of carbon by year.}
    \label{fig:number}
\end{figure}

Figure \ref{fig:number} shows a steady rise in the number of papers and estimates per year. Three papers were published in the 1980s, twenty in 2021. Those 20 papers contained 5,458 estimates of the social carbon. As the literature has grown richer and computers faster, researchers report a wider range of sensitivity analyses.

\begin{figure}
    \centering
    \includegraphics[width = \textwidth]{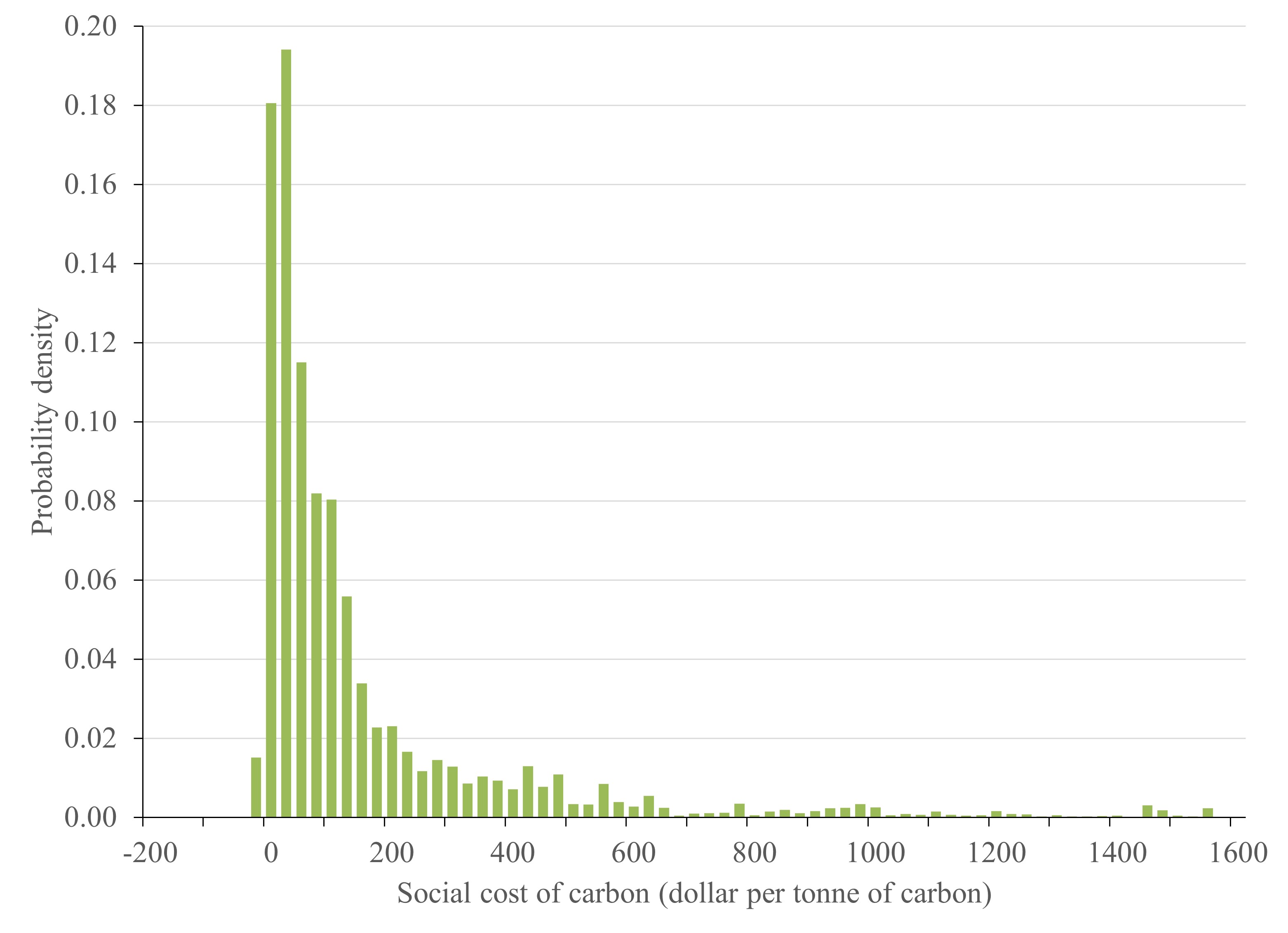}
    \caption{The histogram of published estimates of the social cost of carbon. Estimates are author- and quality-weighted and censored.}
    \label{fig:histo}
\end{figure}

\begin{figure}
    \centering
    \includegraphics[width = \textwidth]{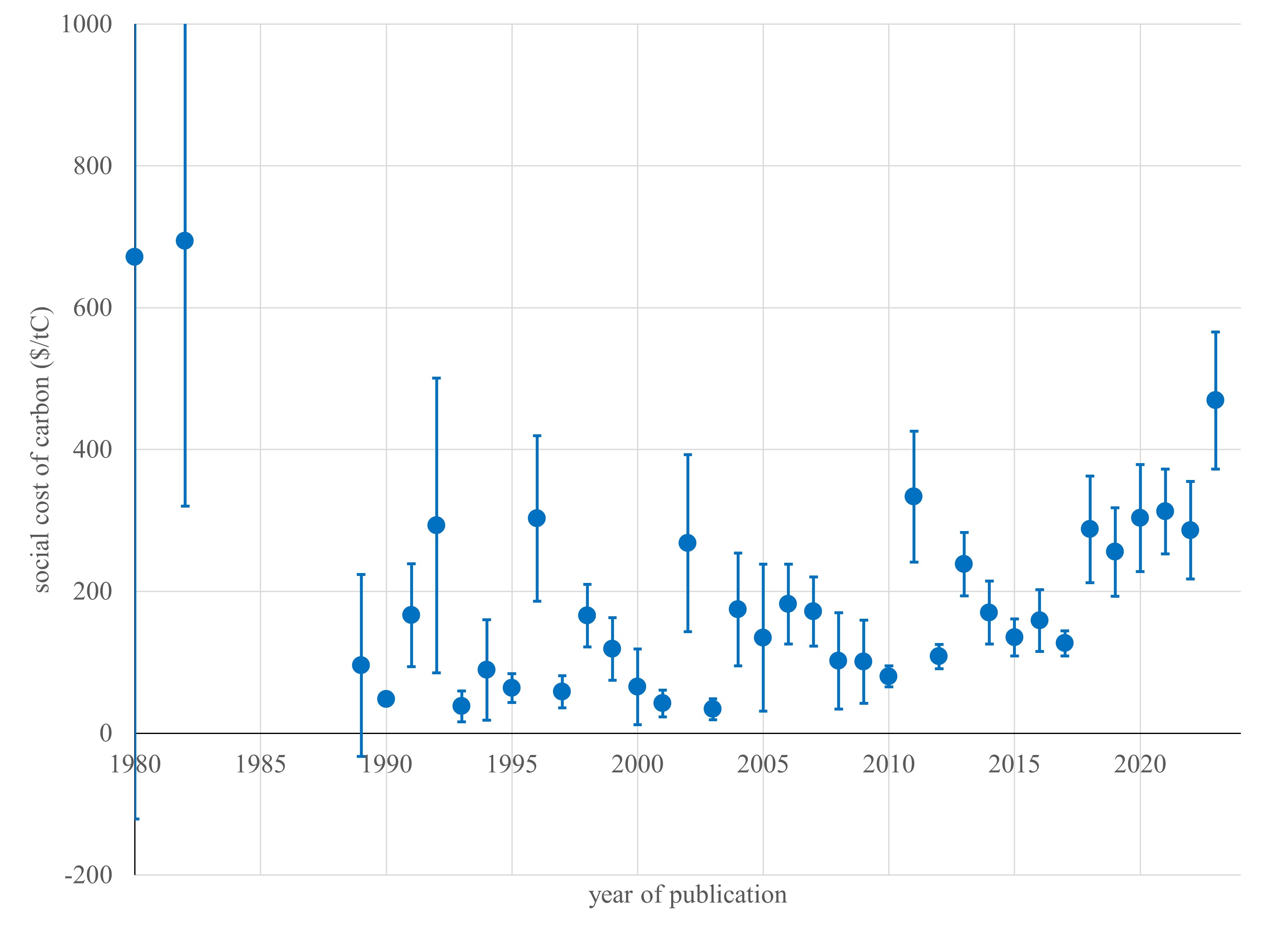}
    \caption{The average social cost of carbon by year of publication. The interval shown is the mean plus and minus the standard deviation. Estimates are author- and quality-weighted and censored.}
    \label{fig:time}
\end{figure}

Estimates of the social cost of carbon started high, fell, and then rose again after 2010; see Figure \ref{fig:time} and \citet{Tol2023NCC}. The mode of all published estimates lies between \$25/tC and \$50/tC. The weighted mean is \$207/tC. The distribution has a pronounced right tail. Only 1.6\% of estimates point to a social \emph{benefit} of carbon. See Figure \ref{fig:histo}.

\begin{figure}
    \centering
    \includegraphics[width = \textwidth]{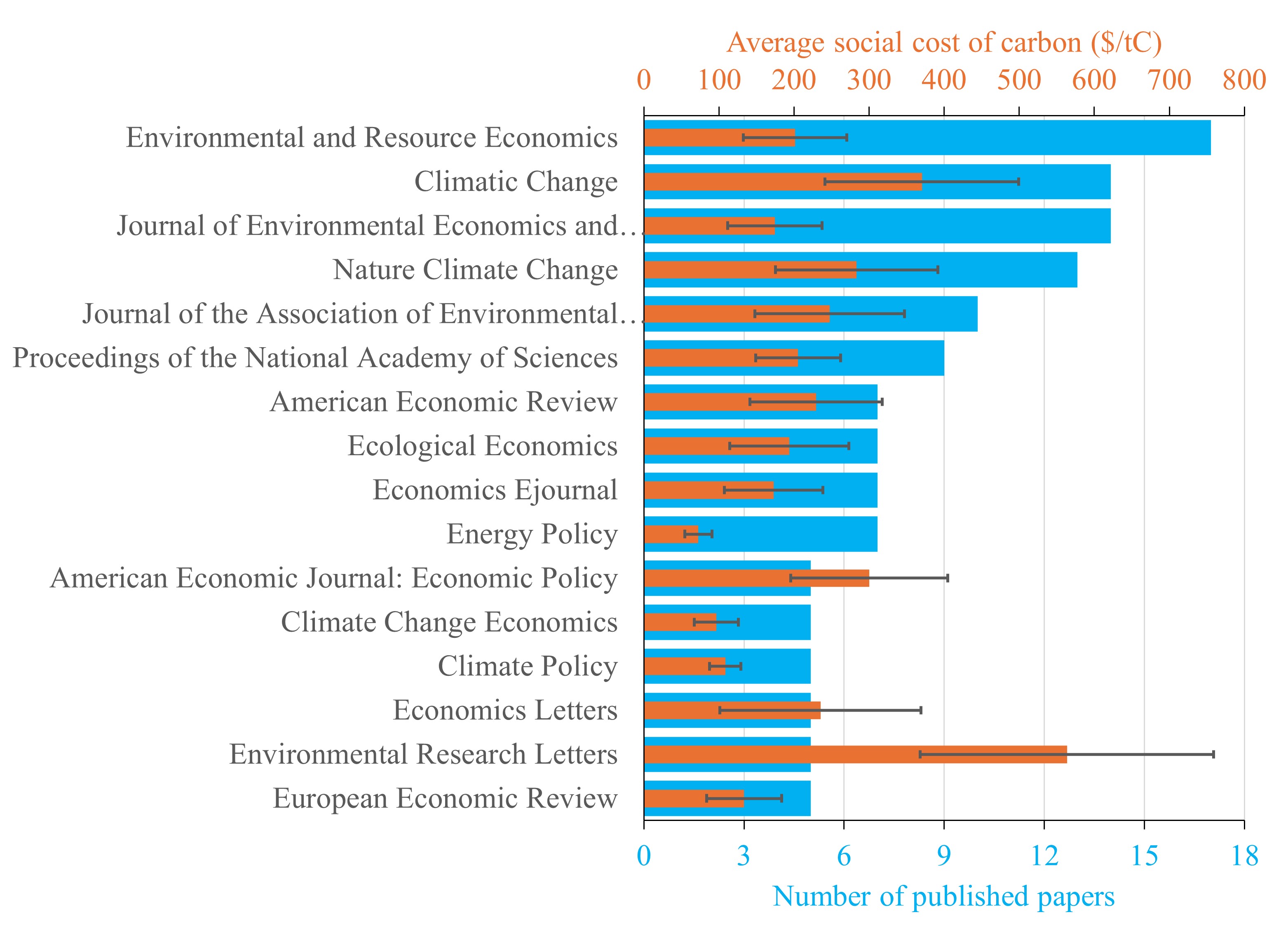}
    \caption{The average social cost of carbon for the sixteen journals that published five or more papers.}
    \label{fig:journal}
\end{figure}

A new feature of the meta-database, Figure \ref{fig:journal} shows the number of papers published per journal, for the sixteen journals that published five or more papers. Figure \ref{fig:journal} also shows the average social cost of carbon for these papers. \textit{Environmental \& Resource Economics} is the most prolific journal with seventeen papers. Estimates of the social cost of carbon are highest, \$564/tC, in \textit{Environmental Research Letters}, and lowest in \textit{Energy Policy}, \$72/tC. Economics journals tend to publish lower estimates than natural science and environmental policy journals. The difference is on average \$86/tC, with a standard error of \$44/tC; $p=0.027$.  

Another new feature of the meta-database, the average paper has 1.03 authors. Estimating the social cost of carbon is a solo activity: Most papers take an existing model and scenarios and tweak a few assumptions. That said, the maximum is 23 authors \citep{Rennert2022}, reflecting the effort needed to build a new integrated assessment model and develop new scenarios.

\begin{figure}
    \centering
    \includegraphics[width = \textwidth]{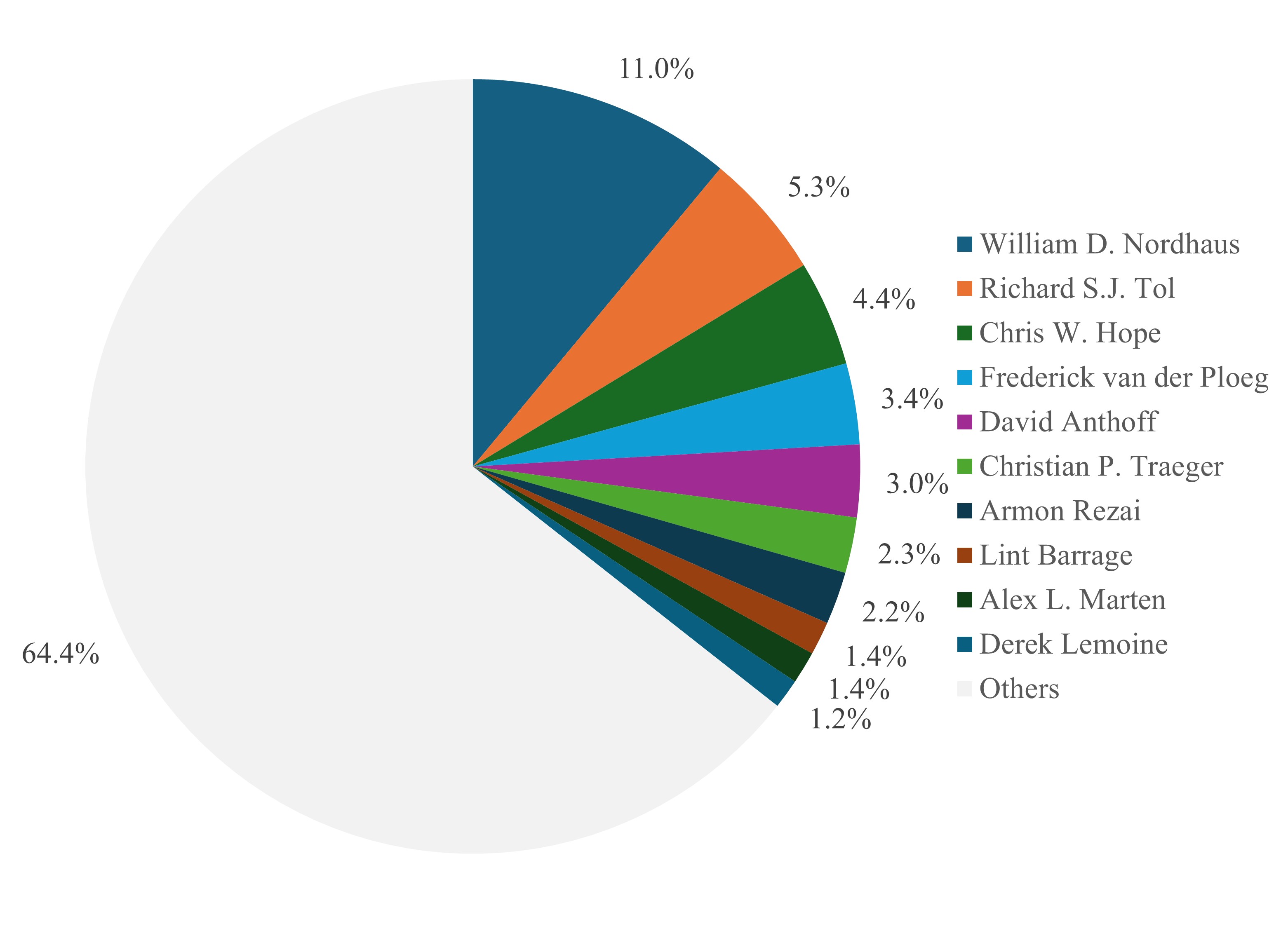}
    \caption{The number of papers on the social cost of carbon by author.}
    \label{fig:author}
\end{figure}

Figure \ref{fig:author} shows the ten most-prolific authors, who together cover almost half of the literature. William D. Nordhaus has made the largest contribution, spanning more than four decades \citep{Nordhaus1980, Barrage2023}.

\begin{figure}
    \centering
    \includegraphics[width = \textwidth]{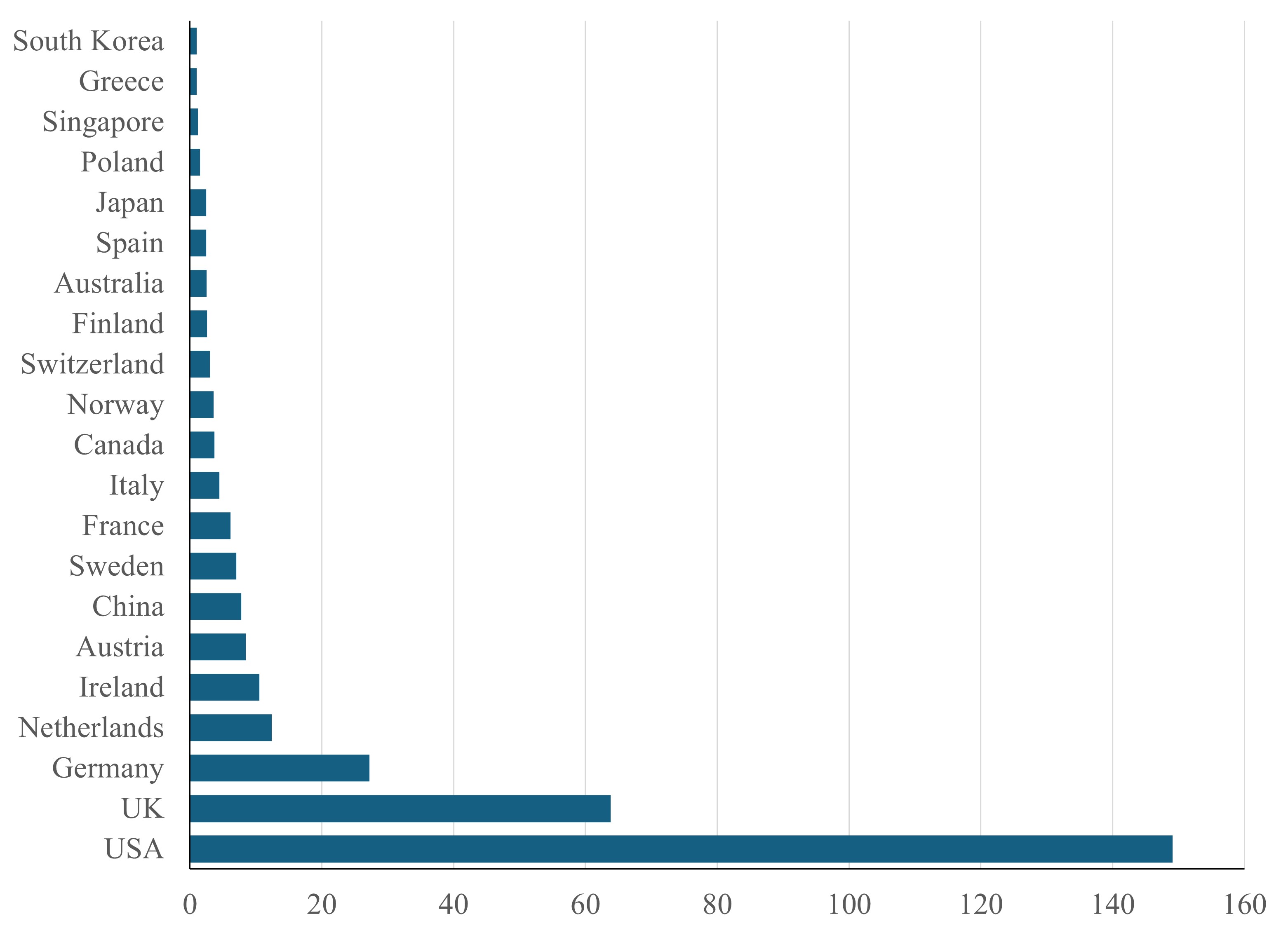}
    \caption{The number of papers on the social cost of carbon by country of affiliation.}
    \label{fig:country}
\end{figure}

Figure \ref{fig:country} shows the country of affiliation of these authors. Almost half of all papers originate in the USA, about a fifth in the UK. Nineteen other countries contributed to this literature. There are few papers from Asia and no papers from Latin America or Africa. This reduces the political legitimacy of estimates of the social cost of carbon. \citet{Dong2024} show that people from the represented countries tend to be more patient and so prefer a higher social cost of carbon.

Figure \ref{fig:coauthor} shows the co-author networks of authors with five or more papers. There are seven such networks, connecting 160 out of 332 authors. The largest network has 116 members, including the top 5 in Figure \ref{fig:author}. There are significant differences between networks. The smallest one, Peck and Teisberg, reports an average social cost of carbon of \$32/tC, significantly smaller than any of the other networks. The results for the Budolfson and Newbold networks are statistically indistinguishable. The largest network, Nordhaus', is in the middle with \$156/tC. The Traeger network finds a higher social cost of carbon on average, but the difference is statistically insignificant. The two remaining networks, Gerlagh and Groom, are again indistinguishable. Indeed, the variation in Groom's network is so large that its average is not significantly different from that of five of the six other networks. Because of these inconclusive results (and because the Nordhaus and Gerlagh networks will merge in the near future), co-author \emph{networks} will not be further considered. I instead focus on individual authors below.

Figure \ref{fig:citation} shows the citation network. The node size is weighted arithmetic incloseness, counting not just the citations of a paper, but also citations of citations, citations of citations of citations, and so on. Weights are inversely proportional to the number of citations in the referring paper. The citation network is dense. There are cycles as people cite early versions. I regressed citation number on publication year and identified the 10\% of papers that have the largest residuals. These papers are most influential for their age. They are named in Figure \ref{fig:citation}. The central-most papers, uncorrected for age, are \citet{Nordhaus1991EJ} and \citet{Nordhaus1994book}. Corrected for age, \citet{Stern2006} comes first, followed by \citet{Nordhaus2008}.

\section{Analysis}
\label{sc:analysis}

\subsection{Social cost of carbon}
Table \ref{tab:base} shows regression results for the social cost of carbon. The base specification includes the pure rate of time preference and the inverse of the elasticity of intertemporal substitution. Higher values for these two parameters imply a higher discount rate \citep{Ramsey1928} and so a lower social cost of carbon. The estimated coefficients have the correct sign. Time preference is always highly significant, utility curvature is significant in the base and most alternative specifications. In the base specification, a one percentage point increase in the pure rate of time preference reduces the social cost of carbon by \$112/tC. An increase in the inverse of the elasticity of intertemporal substitution reduces the social cost of carbon by \$45/tC. These are large changes for modest changes in assumptions.

The base specification also includes the year of publication. Estimates of the social cost of carbon have increased over time but the significance of the upward trend varies with specification. In the base specification, changes in the assumed discount rate explain the observed upward trend (Figure \ref{fig:time}).

Table \ref{tab:base} also shows a number of extended specifications. The second column includes the assumed economic impact of 2.5\celsius{} warming. The social cost of carbon increases with the assumed severity of climate change. If the impact of 2.5\celsius{} warming is 1\% of GDP worse, the social cost of carbon increase by \$20/tC. The shape of the impact function matters too. Commonly used impact functions find a lower social cost of carbon than a smorgasbord of ``other'' or undefined impact functions (the base category)\textemdash but the popular impact functions do not lead to results that differ from other often-used functions. Studies that adopt the findings of econometric estimates of the impact of weather shocks on economic growth \citep{Dell2012, Burke2015} report higher social cost of carbon estimates \citet{Moyer2014, Moore2015}, on average \$265/tC higher.

There is no statistically significant impact of imposing the estimated social cost of carbon as a carbon tax (column 3). The assumed temperature in 2100 does not affect the reported estimate either, nor does the interaction between temperature and the dummy whether a carbon tax is imposed. (Recall that a carbon tax reduces emissions and so global warming and the social cost of carbon.) A large number of studies do not report results for global warming; the sample size shrinks accordingly. This affects the size but not the sign or significance of the welfare parameters. The time trend becomes significant.

The social cost of carbon is lower in papers (co-)authored by Rick van der Ploeg, Richard Tol, or Christian Traeger (column 4). The differences between these authors are not statistically significant.

Figure \ref{fig:journal} shows systematic differences between journals, but these averages do not account for differences in composition. Table \ref{tab:journals} regresses the social cost of carbon on the pure rate of time preference, the inverse of the elasticity of intertemporal substitution, the year of publication, and dummies for the sixteen journals that published five papers or more on the social cost of carbon. Two journals stand out: \textit{Environmental Research Letters} and \textit{Climatic Change} publish estimates of the social cost of carbon that are statistically significantly higher than do other journals.

Table \ref{tab:countries} repeats the exercise for the country of affiliation. There are no significant differences between countries. This is reassuring at the surface: Researchers from different places reach the same conclusion. However, it may be that researchers from unrepresented countries mimic the assumptions of researchers in the countries that dominate this literature (see next section).

\subsection{Publication bias}
Figure \ref{fig:journal} and Table \ref{tab:journals} show systematic differences between journals in the published estimates of the social cost of carbon. Table \ref{tab:journals} tests for differences in the \emph{assumptions} underlying these estimates. This would indicate that certain assumptions are frowned upon by a journal's referees or editors; or that authors believe that certain assumptions would increase their chance of getting published in that journal.

Three journals assume a pure rate of time preference that is significantly higher and one journal lower than the rest. Three other journals have a more curved utility function, and one journal a flatter one. Two journals are more pessimistic about the impact of climate change. However, the two outlier journals identified in Figure \ref{fig:journal} do not use statistically distinct assumptions. Their exceptional estimates are due to a combination of assumptions rather than a single one. There does not seem to be a consistent selection of papers on assumptions. The \textit{Journal of Environmental Economics and Management}, for instance, publishes papers that are more pessimistic on climate change and use a lower inverse of the elasticity of intertemporal substitution (pushing the social cost of carbon up) but use a higher pure rate of time preference (pushing the social cost of carbon down). 

Table \ref{tab:countries} shows no statistically significant differences between countries, with the exception of authors based in Finland and Italy who are more impatient (second column) and risk-averse (third column), respectively, than others. Authors based in Ireland are especially optimistic about the impacts of climate change (fourth column). However, these differences do not lead to differences in the social cost of carbon (first column).

In sum, there is little evidence of \emph{differential} publication bias. While it may be that certain results are more difficult to publish, it is hard to imagine that gatekeeping is consistent across countries and journals. The lack of evidence for differential gatekeeping militates against gatekeeping on results or assumptions.

\subsection{Citation bias}
Table \ref{tab:base}, column 5, shows that estimates of the social cost of carbon are not affected by the social cost of carbon in cited studies. Estimates in studies that are not cited but were cited in cited papers (``aware'') are significant at the 10\% level, as are other non-cited estimates. This suggests that authors preferentially cite lower estimates, perhaps to make their own estimates look more innovative. The time trend become significant in this specification, as the non-cited social cost of carbon falls over time.

Table \ref{tab:journals}, column 5, shows that five journals preferentially refer to papers with higher estimates of the social cost of carbon. One of these journals, \textit{Climatic Change}, preferentially publishes papers with high estimates (column 1); the citation bias may reflect a preference for self-citation. That explanation does not hold for the other four journals. No journal preferentially cites papers with lower estimates of the social cost of carbon.

Table \ref{tab:countries} repeats this exercise for country of affiliation. Authors from Poland selectively cite pessimistic papers. The dummy for Sweden is significant at the 10\% level. As with the journals, there are no significantly negative coefficients. While citation bias may not be widespread, it works in one direction only.

Table \ref{tab:bias} returns to the right-most column of Table \ref{tab:base}. I regress the weighted average of the social cost of carbon in a paper on the average in the cited papers, the average in the papers that were cited in cited papers (``aware'') and in previously published papers that were not cited, controlling for the year of publication. In Table \ref{tab:base}, the effect is significant at the 10\% level. In Table \ref{tab:bias}, the effect of the social cost of carbon reported in the literature on the social cost of carbon is significant at the 5\% level. As above, high estimates of the social cost of carbon are associated with high estimates in non-cited papers. The effect is large: if the social cost of carbon is \$10/tC higher in non-cited (but known) papers, the reported social cost of carbon increases by \$10/tC (\$5/tC). This suggests that people preferentially cite lower estimates to make their own estimate look novel.

This is a paradox. Tables \ref{tab:base} and \ref{tab:bias} show that high estimates are not cited while Table \ref{tab:journals} shows that citing high estimates increases the chance of publication. However, as the latter result only holds for some journals, the two results can be reconciled by authors avoid citing high estimates unless they target the journals identified in Table \ref{tab:journals}.

The assumed pure rate of time preference does not depend on assumptions in previous studies. However, the elasticity of intertemporal substitution and the impact of 2.5\celsius{} do. Particularly, authors that use a high inverse of the elasticity of intertemperal substitution appear to avoid citing studies that do the same. Similarly, studies that have a more optimistic outlook on the impact of 2.5\celsius{} global warming cite similar studies less. 

The rightmost column of Table \ref{tab:bias} confirms the last result. It regresses the number of citations on study characteristics. Papers that are less pessimistic about the impact of climate change, receive fewer citations. These studies report lower estimates of the social cost of carbon (Table \ref{tab:base}) but citation selection is on the total impact of climate change rather than the marginal impact.

\section{Conclusion}
\label{sc:conclude}
I update and extend the meta-analysis of the social cost of climate change. I confirm that the uncertainty about the social cost of carbon is large and right-skewed. The overall mean estimate is \$207/tC. Estimates have trended upward over time. The social cost of carbon is higher if the impact of climate change is worse or if the discount rate is lower. The extension of the meta-database reveals that the literature on the social cost of carbon is dominated by a few networks of co-authors, who are predominantly based in Europe and North America.

There is evidence of publication bias: Some journals prefer to publish higher estimates of the social cost of carbon. This raises the overall average. There is evidence of citation bias too. Authors systematically ignore earlier, higher estimates, which allows them to present their own, higher estimates as novel.

The following caveats apply. Any meta-analysis is only as good as the underlying database. New estimates of the social cost of carbon are published frequently. Methods to estimate the social cost of carbon change, so new fields must be added to the database. Here, I coded broad similarities rather than fine differences between methods and assumptions. Others may take a different approach. The meta-analysis is limited to the mean. Quantile regression, semi- or non-parametric regression, or machine learning methods may lead to somewhat different conclusions.

These things can be deferred to future research. For now, three key findings emerge. (\textit{i}) The social cost of carbon is large and rising. This justifies greenhouse gas emission reduction. (\textit{ii}) The literature on the social cost of carbon suffers from publication or citation bias, exaggerating estimates. (\textit{iii}) The limited geographical spread of researchers may affect the political acceptability of these findings.

\begin{table}[]
    \begin{center}
    \footnotesize
    \caption{Study characteristics and the social cost of carbon}
    \label{tab:base}
    \begin{tabular}{lccccc} \hline
 & SCC & SCC & SCC & SCC & SCC \\ \hline
PRTP & -112.3*** & -123.0*** & -75.77*** & -121.4*** & -108.2*** \\
 & (13.41) & (14.37) & (17.47) & (13.36) & (13.42) \\
EIS & -45.10** & -71.65*** & -100.3*** & -48.38*** & -54.34*** \\
 & (17.57) & (24.37) & (28.90) & (17.61) & (17.55) \\
Year & 2.513 & 0.196 & 7.506*** & 1.849 & 6.973*** \\
 & (1.581) & (1.758) & (2.194) & (1.728) & (2.065) \\
Impact &  & -19.97*** &  &  &  \\
 &  & (4.891) &  &  &  \\
Linear &  & -207.8** &  &  &  \\
 &  & (98.56) &  &  &  \\
Quadratic &  & -65.00* &  &  &  \\
 &  & (34.25) &  &  &  \\
Parabola &  & -110.6* &  &  &  \\
 &  & (60.88) &  &  &  \\
Weitzman &  & -180.9** &  &  &  \\
 &  & (83.76) &  &  &  \\
\textsc{fund} &  & -101.3** &  &  &  \\
 &  & (45.36) &  &  &  \\
Growth &  & 265.4*** &  &  &  \\
 &  & (98.03) &  &  &  \\
Tax &  &  & 178.0 &  &  \\
 &  &  & (162.3) &  &  \\
$T_{2100}$ &  &  & -6.923 &  &  \\
 &  &  & (40.83) &  &  \\
$\textrm{Tax}\times T_{2100}$ &  &  & -53.43 &  &  \\
 &  &  & (49.21) &  &  \\
Hope &  &  &  & -10.46 &  \\
 &  &  &  & (56.52) &  \\
Nordhaus &  &  &  & -58.43 &  \\
 &  &  &  & (43.88) &  \\
van der Ploeg &  &  &  & -176.1*** &  \\
 &  &  &  & (49.08) &  \\
Tol &  &  &  & -196.2*** &  \\
 &  &  &  & (41.14) &  \\
Traeger &  &  &  & -131.7** &  \\
 &  &  &  & (56.53) &  \\
Cited SCC &  &  &  &  & 0.140 \\
 &  &  &  &  & (0.105) \\
Not cited &  &  &  &  & 0.930* \\
 &  &  &  &  & (0.516) \\
Aware &  &  &  &  & 0.359* \\
 &  &  &  &  & (0.207) \\
Constant & -4,635 & 103.0 & -14,699*** & -3,227 & -13,875*** \\
 & (3,185) & (3,538) & (4,403) & (3,483) & (4,213) \\ \hline
Observations & 787 & 706 & 393 & 787 & 781 \\
 R-squared & 0.097 & 0.163 & 0.139 & 0.136 & 0.109 \\ \hline
\multicolumn{6}{c}{ Standard errors in parentheses; *** p$<$0.01, ** p$<$0.05, * p$<$0.1} \\
\end{tabular}

    \end{center}
\end{table}

\begin{table}[]
    \begin{center}
    \footnotesize
    \caption{Differences between journals}
    \label{tab:journals}
    \begin{tabular}{lccccc} \hline
 & SCC & PRTP & EIS & impact & Cited SCC \\ \hline
PRTP & -115.1*** &   &  &  & \\
 & (13.56) &   &  &  & \\
EIS & -45.15** &   &  &  & \\
 & (17.71) &   &  &  & \\
Year & 2.382 & -0.0236*** & 0.00111 & -0.0470*** & -6.587*** \\
 & (1.699) & (0.00447) & (0.00316) & (0.0112) & (0.534)\\
European Econ Rev & -75.44 & 0.153 & -0.393** & 0.318 & 13.56 \\
 & (86.65) & (0.234) & (0.176) & (0.894) & (33.49) \\
Env Research Let & 333.5*** & -0.195 & -0.115 & 1.063 & 4.184  \\
 & (93.02) & (0.252) & (0.190) & (0.789) & (34.77)  \\
Econ Let & 58.51 & -0.159 & 0.481*** & -1.552** & -46.22 \\
 & (85.42) & (0.231) & (0.174) & (0.649) & (33.10)  \\
Climate Pol & -52.72 & 0.0968 & -0.285 & 0.453 & -41.80 \\
 & (94.83) & (0.257) & (0.188) & (0.777) & (32.18)\\
Climate Change Econ & -95.43 & -0.268 & 0.0996 & 1.076 & -1.143 \\
 & (114.6) & (0.311) & (0.234) & (0.730) & (33.16) \\
AEJ Econ Pol & 51.44 & 0.213 & 0.329 & -1.152 & 29.58 \\
 & (99.59) & (0.270) & (0.203) & (0.712) & (36.24) \\
Energy Pol & 74.71 & 0.860*** & -0.286 & -0.735 & 66.58**\\
 & (116.5) & (0.275) & (0.236) & (0.571) & (28.11) \\
Econ Ejournal & -126.2 & -0.663*** & -0.156 & 0.508 & -35.90  \\
 & (87.81) & (0.237) & (0.178) & (0.538) & (27.40)  \\
Ecol Econ & 69.02 & 0.607*** & -0.0952 & 0.738 & 41.30  \\
 & (71.36) & (0.192) & (0.145) & (0.540) & (27.48) \\
American Econ Rev & 67.16 & 0.299 & 0.217 & -0.202 & 106.3***  \\
 & (87.77) & (0.238) & (0.170) & (0.636) & (32.24)  \\
Proc Nat Acad Science & 44.02 & 0.0733 & 0.373*** & -0.0170 & 21.95  \\
 & (67.64) & (0.183) & (0.132) & (0.589) & (24.98)  \\
J Assoc Env Res Econ & 27.22 & -0.116 & 0.0312 & 0.965* & 67.61***  \\
 & (79.48) & (0.216) & (0.147) & (0.538) & (24.78)  \\
Nature Climate Change & -33.18 & 0.159 & 0.255** & 0.0854 & 30.16  \\
 & (62.09) & (0.168) & (0.124) & (0.461) & (22.42)  \\
J Env Econ Mgt & -4.130 & 0.377** & -0.217* & -1.705*** & 82.05***  \\
 & (70.75) & (0.182) & (0.121) & (0.460) & (20.76) \\
Climatic Change & 227.3*** & 0.107 & -0.0527 & -0.358 & 75.84***  \\
 & (57.50) & (0.156) & (0.113) & (0.406) & (19.93)  \\
Env Res Econ & -38.26 & 0.242 & 0.143 & -0.140 & 45.71** \\
 & (57.89) & (0.157) & (0.114) & (0.423) & (20.16) \\
Constant & -4,387 & 48.96*** & -0.949 & 92.96*** & 13,422***  \\
 & (3,420) & (8.988) & (6.360) & (22.53) & (1,073) \\ \hline
Observations & 787 & 807 & 871 & 911 & 1,078  \\
 R-squared & 0.139 & 0.086 & 0.047 & 0.057 & 0.175  \\ \hline
\multicolumn{6}{c}{ Standard errors in parentheses; *** p$<$0.01, ** p$<$0.05, * p$<$0.1} \\
\end{tabular}

    \end{center}
    {\footnotesize Column 1 regresses the social cost of carbon on dummies for the sixteen journals that published five papers or more. Columns 2-5 regress key assumptions on said dummies. All results are quality-weighted and censored.}
\end{table}

\begin{table}[]
    \begin{center}
    \footnotesize
    \caption{Differences between countries}
    \label{tab:countries}
    \begin{tabular}{lccccc} \hline
& SCC & PRTP & EIS & Impact & Cited SCC \\ \hline
 &  &  &  &  &  \\
PRTP & -115.7*** &  &  &  &  \\
 & (15.74) &  &  &  &  \\
EIS & -28.58 &  &  &  &  \\
 & (19.82) &  &  &  &  \\
Year & 2.030 & -0.0261*** & 1.79e-05 & -0.0221* & -7.551*** \\
 & (1.901) & (0.00517) & (0.00374) & (0.0113) & (0.645) \\
Austria & 332.5 & -0.197 & 0.236 & -2.361 & 26.95 \\
 & (230.8) & (0.647) & (0.499) & (1.461) & (90.91) \\
China & 179.0 & 0.482 & -0.451 & -0.443 & 84.63 \\
 & (208.8) & (0.577) & (0.446) & (1.416) & (76.68) \\
Finland & 354.9 & 1.526** & -1.79e-05 & 1.196 & -0.466 \\
 & (263.0) & (0.735) & (0.567) & (1.686) & (104.9) \\
France & 252.6 & -0.338 & -0.0695 & -2.065 & -13.19 \\
 & (214.6) & (0.580) & (0.451) & (1.299) & (78.41) \\
Germany & 254.3 & -0.333 & -0.308 & 1.320 & 16.12 \\
 & (190.8) & (0.535) & (0.412) & (1.167) & (72.05) \\
Ireland & -44.83 & -0.381 & -0.257 & 3.108** & -59.41 \\
 & (204.2) & (0.572) & (0.441) & (1.275) & (78.53) \\
Italy & 57.41 & 0.0542 & 1.188** & 1.098 & 76.05 \\
 & (226.4) & (0.632) & (0.487) & (1.421) & (88.41) \\
Japan & -22.75 & -0.0809 & 0.530 & 0.943 & 29.66 \\
 & (261.8) & (0.734) & (0.566) & (1.684) & (98.63) \\
Netherlands & 312.3 & 0.657 & -0.413 & -1.135 & 100.3 \\
 & (260.8) & (0.731) & (0.494) & (1.410) & (83.23) \\
Norway & 142.8 & -0.0494 & -0.198 & 0.821 & 147.9* \\
 & (215.6) & (0.605) & (0.466) & (1.350) & (83.99) \\
Poland & 143.3 & 0.157 & -0.0308 & 1.126 & 352.7*** \\
 & (263.4) & (0.739) & (0.570) & (1.696) & (105.5) \\
South Korea &  &  &  &  & 81.76 \\
 &  &  &  &  & (107.7) \\
Sweden & 334.8 & -0.838 & -0.298 & 0.678 & 152.1* \\
 & (206.0) & (0.576) & (0.444) & (1.276) & (79.33) \\
Switzerland & -24.60 & 0.0523 & -0.450 &  & 82.12 \\
 & (281.3) & (0.789) & (0.608) &  & (113.4) \\
UK & 151.2 & -0.145 & -0.461 & 1.034 & 25.86 \\
 & (188.5) & (0.528) & (0.406) & (1.143) & (70.85) \\
USA & 117.6 & 0.0534 & -0.0447 & 0.873 & 33.63 \\
 & (185.9) & (0.521) & (0.402) & (1.129) & (70.03) \\
Constant & -3,805 & 54.23*** & 1.414 & 41.81* & 15,339*** \\
 & (3,845) & (10.45) & (7.545) & (22.79) & (1,304) \\ \hline
Observations & 556 & 571 & 615 & 637 & 775 \\
 R-squared & 0.154 & 0.102 & 0.087 & 0.095 & 0.191 \\ \hline
\multicolumn{6}{c}{ Standard errors in parentheses; *** p$<$0.01, ** p$<$0.05, * p$<$0.1} \\
\end{tabular}

    \end{center}
    {\footnotesize Column 1 regresses the social cost of carbon on dummies for the countries of affiliation. The base category is multi-country. Columns 2-5 regress key assumptions on said dummies. All results are quality-weighted and censored.}
\end{table}

\begin{table}[]
    \begin{center}
    \footnotesize
    \caption{Evidence of citation bias}
    \label{tab:bias}
    \begin{tabular}{lccccc} \hline
 & SCC & PRTP & EIS & Benchmark & Citations \\ \hline
Cited & -0.0223 & 0.00877 &  0.0956 & 0.0440 &   \\
 & (0.115) & (0.118) & (0.106) & (0.117) &   \\
Aware & 0.462** & -0.267 & -0.241* & 0.0161 &  \\
 & (0.205) &  (0.232) & (0.129) & (0.289)  &  \\
Not & 0.956** & 0.931 &   1.146*** & 1.636** &  \\
 & (0.461) & (0.577) &   (0.172) & (0.717)  &  \\
Year & 5.457** & -0.0102 & 0.0125* & -0.0117 & -0.0544*** \\
 &  (2.191) & (0.0134) & (0.00697) & (0.0131) & (0.0125) \\
SCC & &  &  &  & -0.000766  \\
 & &  &  &  & (0.000511) \\
PRTP &  &  &  &  &  -0.162 \\
 &  &  &  &  &  (0.133)  \\
EIS &  &  &  &  &  -0.282  \\
 &  &  &  &  &   (0.208)  \\
Benchmark &  &  &  &  &  -0.105**  \\
 &  &  &  &  &  (0.0508)  \\
$\ln \alpha$   &  &  &  &  & 0.520*** \\
   &  &  &  &  & (0.117) \\
Constant &  -11,052** & 20.88 & -25.21* & 24.33 & 112.1*** \\
 & (4,456) & (27.81) & (14.22) & (26.10) & (25.27) \\
Observations & 317 & 190 & 230 & 247 & 178 \\
 R-squared &0.034 & 0.094 & 0.244 & 0.034 &  \\ \hline
\multicolumn{6}{c}{ Standard errors in parentheses} \\
\multicolumn{6}{c}{ *** p$<$0.01, ** p$<$0.05, * p$<$0.1} \\
\end{tabular}
    \end{center}
    {\footnotesize Columns 1-4 regress the social cost of carbon, pure rate of time preference, inverse of the elasticity of intertemporal substitution and impact of 2.5\celsius{} warming, average over all estimates in a paper, on the same indicators averaged over all papers cited, over papers not cited but cited in the cited papers (``aware''), and over other papers not cited. Column 5 uses a Negative Binomial regression of the number of citations on the characteristics of the cited paper; the estimate for $\ln \alpha$ indicates that this specification outperforms a Poisson regression.}
\end{table}

\newpage\bibliography{master}

\begin{thebibliography}{28}
\providecommand{\natexlab}[1]{#1}

\bibitem[{Anthoff and Tol(2022)}]{Anthoff2022}
Anthoff, David, and Richard S.~J. Tol. 2022.
\newblock Testing the dismal theorem.
\newblock \emph{Journal of the Association of Environmental and Resource Economists} 9~(5): 885--920.

\bibitem[{Barrage and Nordhaus(2023)}]{Barrage2023}
Barrage, Lint, and William~D. Nordhaus. 2023.
\newblock Policies, projections, and the social cost of carbon: Results from the {DICE-2023} model.
\newblock Cowles Foundation Discussion Papers 2363, Cowles Foundation for Research in Economics, Yale University.

\bibitem[{Burke et~al.(2015)Burke, Hsiang, and Miguel}]{Burke2015}
Burke, Marshall, Solomon~M. Hsiang, and Edward Miguel. 2015.
\newblock Global non-linear effect of temperature on economic production.
\newblock \emph{Nature} 527~(7577): 235--239.

\bibitem[{Dell et~al.(2012)Dell, Jones, and Olken}]{Dell2012}
Dell, Melissa, Benjamin~F. Jones, and Benjamin~A. Olken. 2012.
\newblock Temperature shocks and economic growth: Evidence from the last half century.
\newblock \emph{American Economic Journal: Macroeconomics} 4~(3): 66--95.

\bibitem[{Dong et~al.(2024)Dong, Tol, and Wang}]{Dong2024}
Dong, Jinchi, Richard S.~J. Tol, and Fangzhi Wang. 2024.
\newblock Towards a social cost of carbon with national characteristics.
\newblock \emph{Economics Letters} .

\bibitem[{Greenstone et~al.(2013)Greenstone, Kopits, and Wolverton}]{Greenstone2013}
Greenstone, M., E.~Kopits, and A.~Wolverton. 2013.
\newblock Developing a social cost of carbon for us regulatory analysis: A methodology and interpretation.
\newblock \emph{Review of Environmental Economics and Policy} 7~(1): 23--46.

\bibitem[{Havránek et~al.(2015)Havránek, Irsova, Janda, and Zilberman}]{Havranek2015}
Havránek, Tomáš, Zuzana Irsova, Karel Janda, and David Zilberman. 2015.
\newblock Selective reporting and the social cost of carbon.
\newblock \emph{Energy Economics} 51: 394 -- 406.

\bibitem[{Jannot et~al.(2013)Jannot, Agoritsas, Gayet-Ageron, and Perneger}]{Jannot2013}
Jannot, Anne-Sophie, Thomas Agoritsas, Angèle Gayet-Ageron, and Thomas~V. Perneger. 2013.
\newblock Citation bias favoring statistically significant studies was present in medical research.
\newblock \emph{Journal of Clinical Epidemiology} 66~(3): 296 – 301.

\bibitem[{Moore and Diaz(2015)}]{Moore2015}
Moore, Frances~C., and Delavane~B. Diaz. 2015.
\newblock Temperature impacts on economic growth warrant stringent mitigation policy.
\newblock \emph{Nature Climate Change} 5~(2): 127--131.

\bibitem[{Moyer et~al.(2014)Moyer, Woolley, Matteson, Glotter, and Weisbach}]{Moyer2014}
Moyer, Elisabeth~J., Mark~D. Woolley, Nathan~J. Matteson, Michael~J. Glotter, and David~A. Weisbach. 2014.
\newblock Climate impacts on economic growth as drivers of uncertainty in the social cost of carbon.
\newblock \emph{Journal of Legal Studies} 43~(2): 401--425.

\bibitem[{NAS(2017)}]{NAS2017}
NAS. 2017.
\newblock \emph{Valuing climate damages: Updating estimation of the social cost of carbon dioxide}.
\newblock Washington, D.C.: National Academies of Sciences, Engineering, and Medicine.

\bibitem[{Nieminen et~al.(2007)Nieminen, Rucker, Miettunen, Carpenter, and Schumacher}]{Nieminen2007}
Nieminen, Pentti, Gerta Rucker, Jouko Miettunen, James Carpenter, and Martin Schumacher. 2007.
\newblock Statistically significant papers in psychiatry were cited more often than others.
\newblock \emph{Journal of Clinical Epidemiology} 60~(9): 939 – 946.

\bibitem[{Nordhaus(1980)}]{Nordhaus1980}
Nordhaus, William~D. 1980.
\newblock Thinking about carbon dioxide: Theoretical and empirical aspects of optimal control strategies.
\newblock Discussion Paper 565, Cowles Foundation for Research in Economics.

\bibitem[{Nordhaus(1991)}]{Nordhaus1991EJ}
Nordhaus, William~D. 1991.
\newblock To slow or not to slow: The economics of the greenhouse effect.
\newblock \emph{Economic Journal} 101~(444): 920--937.

\bibitem[{Nordhaus(1994)}]{Nordhaus1994book}
Nordhaus, William~D. 1994.
\newblock \emph{Managing the Global Commons: The Economics of Climate Change}.
\newblock Cambridge: The MIT Press.

\bibitem[{Nordhaus(2008)}]{Nordhaus2008}
Nordhaus, William~D. 2008.
\newblock \emph{A Question of Balance\textemdash Weighing the Options on Global Warming Policies}.
\newblock New Haven: Yale University Press.

\bibitem[{Pezzey(2019)}]{Pezzey2019}
Pezzey, John C.~V. 2019.
\newblock Why the social cost of carbon will always be disputed.
\newblock \emph{WIREs Climate Change} 10~(1): e558.

\bibitem[{Pizer et~al.(2014)Pizer, Adler, Aldy, Anthoff, Cropper, Gillingham, Greenstone et~al.}]{Pizer2014}
Pizer, William~A., M.~Adler, J.~Aldy, David Anthoff, Maureen Cropper, K.~Gillingham, M.~Greenstone, et~al. 2014.
\newblock Using and improving the social cost of carbon.
\newblock \emph{Science} 346~(6214): 1189--1190.

\bibitem[{Radicchi and Castellano(2012)}]{Radicchi2012}
Radicchi, Filippo, and Claudio Castellano. 2012.
\newblock A reverse engineering approach to the suppression of citation biases reveals universal properties of citation distributions.
\newblock \emph{PLoS ONE} 7~(3).

\bibitem[{Ramsey(1928)}]{Ramsey1928}
Ramsey, F.~P. 1928.
\newblock A mathematical theory of saving.
\newblock \emph{The Economic Journal} 38~(152): 543--559.

\bibitem[{Rennert et~al.(2022)Rennert, Errickson, Prest, Rennels, Newell, Pizer, Kingdon et~al.}]{Rennert2022}
Rennert, Kevin, Frank Errickson, Brian~C. Prest, Lisa Rennels, Richard~G. Newell, William~A. Pizer, Cora Kingdon, et~al. 2022.
\newblock Comprehensive evidence implies a higher social cost of {CO}\textsubscript{2}.
\newblock \emph{Nature} 610: 687–692.

\bibitem[{Stern et~al.(2006)Stern, Peters, Bakhski, Bowen, Cameron, Catovsky, Crane et~al.}]{Stern2006}
Stern, Nicholas~H., Siobhan Peters, Vicky Bakhski, Alex Bowen, Catherine Cameron, Sebastian Catovsky, Diane Crane, et~al. 2006.
\newblock \emph{Stern Review: The Economics of Climate Change}.
\newblock London: HM Treasury.

\bibitem[{Tol(2005)}]{Tol2005}
Tol, Richard S.~J. 2005.
\newblock The marginal damage costs of carbon dioxide emissions: an assessment of the uncertainties.
\newblock \emph{Energy Policy} 33: 2064--2074.

\bibitem[{Tol(2011)}]{Tol2011}
Tol, Richard S.~J. 2011.
\newblock The social cost of carbon.
\newblock \emph{Annual Review of Resource Economics} 3: 419--443.

\bibitem[{Tol(2018)}]{Tol2018REEP}
Tol, Richard S.~J. 2018.
\newblock The economic impacts of climate change.
\newblock \emph{Review of Environmental Economics and Policy} 12~(1): 4--25.

\bibitem[{Tol(2023)}]{Tol2023NCC}
Tol, Richard S.~J. 2023.
\newblock Social cost of carbon estimates have increased over time.
\newblock \emph{Nature Climate Change} 13: 532--536.

\bibitem[{Tol(2022)}]{Tol2023meta}
Tol, Richard~S.J. 2022.
\newblock {A meta-analysis of the total economic impact of climate change}.
\newblock Working Paper Series 0422, Department of Economics, University of Sussex Business School.

\bibitem[{Wang et~al.(2019)Wang, Deng, Zhou, and Yu}]{Wang2019}
Wang, Pei, Xiangzheng Deng, Huimin Zhou, and Shangkun Yu. 2019.
\newblock Estimates of the social cost of carbon: A review based on meta-analysis.
\newblock \emph{Journal of Cleaner Production} 209: 1494 -- 1507.

\end{thebibliography}

\begin{landscape}
\begin{figure}
    \centering
    \includegraphics[width = 1.2\textwidth]{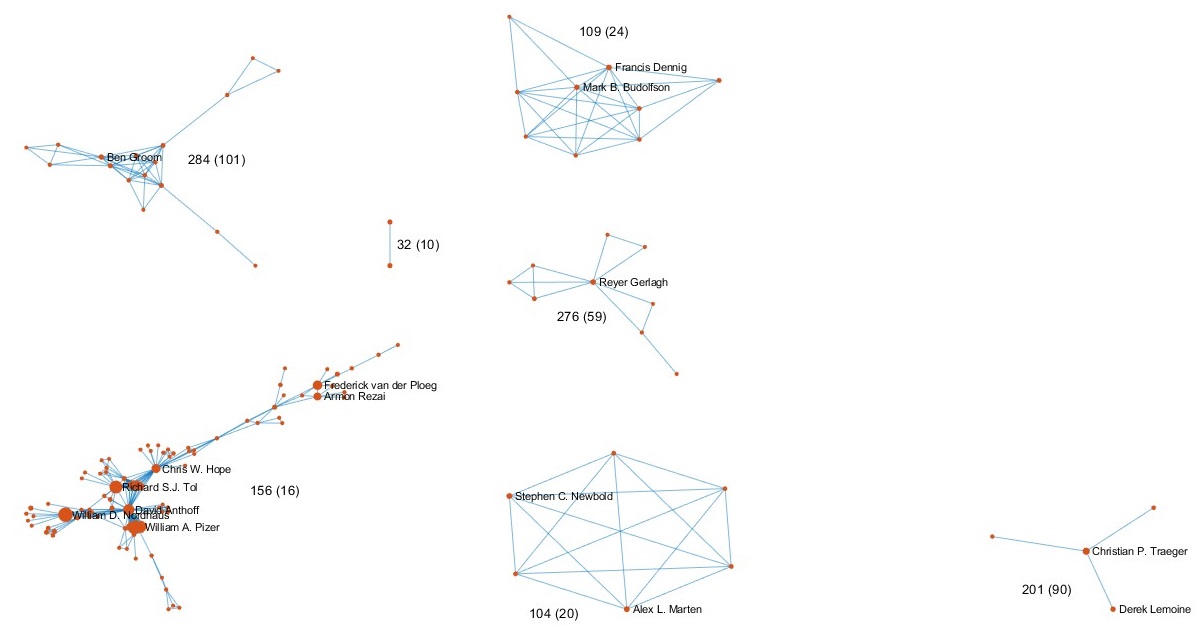}
    \caption{The seven most prolific co-author networks. Node size is the number of co-authored papers. Authors of six papers or more are named. The numbers are the average social cost of carbon and its standard error for papers published by authors in the network.}
    \label{fig:coauthor}
\end{figure}

\begin{figure}
    \centering
    \includegraphics[width = 1.2\textwidth]{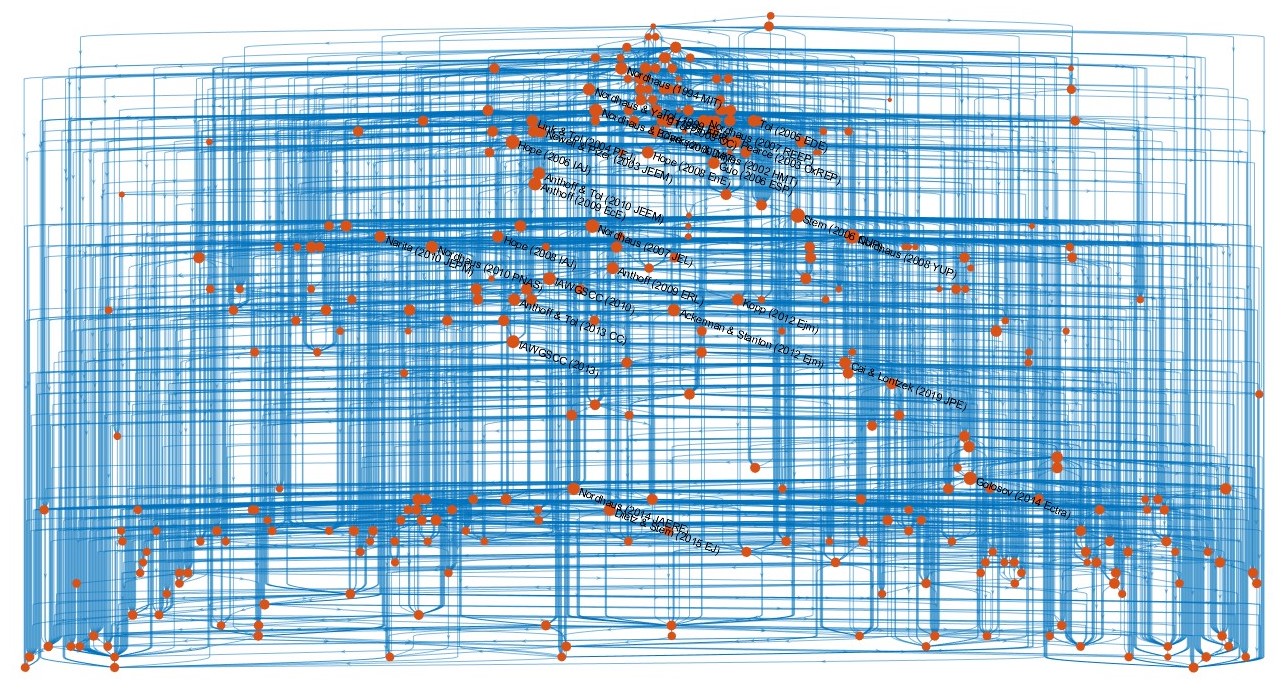}
    \caption{The citation network of papers on the social cost of carbon. Node size is incloseness corrected for number of citation and age.}
    \label{fig:citation}
\end{figure}
\end{landscape}

\end{document}